\documentclass{article}   

\usepackage{bm}
\usepackage{amsmath,amssymb}
\usepackage{graphicx}
\usepackage{listings}

\title{A Visualization Method of Four Dimensional Polytopes by Oval Display of Parallel Hyperplane Slices}

\author{Akira Kageyama\footnote{\texttt{kage@port.kobe-u.ac.jp} } \\[1em]
              Department of Computational Science,\\[0.5em]
               Kobe University, Kobe 657-8501, Japan
}
\date{}

\begin{document}

\maketitle

\abstract{
A method to visualize polytopes in a four dimensional euclidian space $(x,y,z,w)$ is proposed.
A polytope is sliced by multiple hyperplanes that are parallel each other and separated by uniform intervals.
Since the hyperplanes are perpendicular to the $w$ axis,
the resulting multiple slices appear in the three-dimensional $(x,y,z)$ space and they are shown by the standard computer graphics.
The polytope is rotated extrinsically in the four dimensional space by means of a simple input method based on keyboard typings.
The multiple slices are placed on a parabola curve in the three-dimensional world coordinates.
The slices in a view window form an oval appearance.
Both the simple and the double rotations in the four dimensional space are applied to the polytope.
All slices synchronously change their shapes when a rotation is applied to the polytope.
The compact display in the oval of many slices with the help of quick rotations facilitate a grasp of the four dimensional configuration of the polytope.
}

\section{Introduction}

In famous book ``Flatland'', 
an inhabitant called Square living in a flat 2-dimensional (2-D) space meets
a stranger, Sphere, living in a 3-D space~\cite{Abbott:1992}.
Mr.~Square in fact meets Sphere's cross section, which is a circle in Flatland.
We apply in this paper
Mr.~Square's experience to comprehend polytopes in a 4-D euclidian space $(x,y,z,w)$.

One of the earliest studies on visualization of a 4-D polytope is by Noll in 1968~\cite{noll1968computer},
in which a hypercube is visualized by a perspective projection in 4-D.
The resulting 3-D objects, a cube-within-a-cube, is drawn with the wireframe graphics.
When the hypercube is rotated in 4-D,
an intriguing turning-inside-out motion is observed.
The inner cube grows until it becomes the outer cube while the outer cube shrinks until it becomes the inner cube.
This motion was filmed as an animation.

%
%
%

Since the work of Noll, the projection method is the major approach to the visualization of 4-D polytopes.
The approach is not restricted to the computer graphics.
Physical models are built for the projected 4-D regular polytopes with ball-and-sticks~\cite{sequin20023d}.
Mathematics of the 4-D regular polytopes and their graphics with Mathematica is found in~\cite{Sullivan:TheMathematicaJournal:1991}.

MeshView~\cite{hanson1999meshview} is a program to draw 4-D objects 
by perspective and orthogonal projections in 4-D with a special emphasis on a 4-D rolling ball interface.
Input devices for 4-D rotations and translations are studied in detail in~\cite{banks1992interactive}.
More advanced control of 4-D transformations is developed by making use of a head tracking system~\cite{sakai2011four},
in which a 3-D motion of the user's head is linked to a view point motion in 4-D space.
Various rendering methods have been studied for projected objects in 4-D~\cite{hanson1991visualizing,hoffmann1991some,Hanson:IeeeXplore:1993,Chu:VCG:2009}
Some of them are applied to data visualization of 4-D scalar field~\cite{Hanson:FourDimScalar:1992,woodring2003high}.

A very different approach to the visualization of 4-D objects is slicings by a hyperplane in 4-D.
In contrast to the projection method,
studies on the slicing approach is rather sparse.
Hausmann and Seidel developed a program to visualize the six regular polytopes in 4-D~\cite{Hausmann:ComputerGraphicsForum:1994}.
Although their program supports both the projection and the hyperplane slicing approaches, the slice is very restrictive: 
The hyperplane is always along one of the symmetry axes of each polytope.
A visualization of 4-D object by the slicing with an arbitrary hyperplane was investigated by Woodring et al.~\cite{woodring2003high}.
Since the purpose of their study is to visualize time varying, three-dimensional scalar field,
the $w$ axis corresponds to time
and the 3-D slices were visualized by the 3-D volume rendering.

We propose in this paper a 4-D visualization method of polytopes in a 4-D $(x,y,z,w)$ space by taking multiple slices by parallel hyperplanes.
The basic idea and procedure are as follows:
The slicing hyperplanes are intentionally restricted to $w=0$ and its parallels. ($w=\text{constants.}$)
Instead of considering every hyperplanes, i.e.,
$c_x x + c_y y + c_z z + c_w w = c_0$, where $c_x$ etc.~are constants, 
we rotate the polytope in 4-D.
The rotation is designed to be applied extrinsically, i.e., the user observe the object
under a rest frame and apply the 4-D rotations under the fixed coordinates.
The extrinsic rotation with fixed slicing hyperplanes enables the user to keep track of 
the sense of orientation in 4-D space.
The slicing hyperplanes are aligned along the $w$ axis with uniform intervals.
It is like the computed tomography applied to 4-D.
The resulting multiple slices, which are 3-D objects, are presented in a single window
by the standard 3-D computer graphics.
An interactive 4-D rotation invoked by the user causes synchronized deformations of all the 3-D slices.

%
%
%
One of the most interesting applications of the 4-D visualization is the scientific visualization.
The computational fluid dynamics (CFD), for example, produces field data that are defined in 3-D spatial and 1-D temporal dimensions.
When a characteristic spatial structure in a fluid, such as a localized vortex line, exhibits an intriguing dynamical deformation in time,
researchers would try to understand the vortex dynamics through the visualization of the data fields in the 4-D space.
An approach to the 4-D data visualization is the projection-based method that is represented by Noll's work,
in which the field data are integrated in a specific direction to reduce the dimension.
This approach is appropriate for grasping a whole picture of the 4-D data, or for observing macroscopic structures of the spatio-temporal coherency of the data.
In contrast to that, our approach to the 4-D data visualization proposed in this paper
is appropriate for observing microscopic structures of the spatio-temporal coherency in detail,
because the numerical data on specific slice planes are explicitly visualized in our method.

\section{A Pentachoron and its Hyperplane Slices}

Although the method proposed in this paper is a general framework applicable to any 4-D object,
we use it for the visualization of 4-D polytopes, especially a regular pentachoron.
\begin{figure}[t]
  \centering
  \includegraphics[%
      height=0.7\textheight,%
       width=0.7\hsize,keepaspectratio]%
        {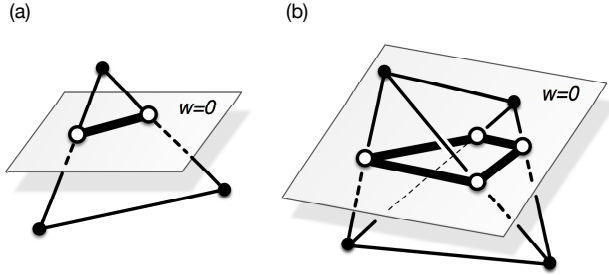}
  \caption{(a) A cross section of a triangle in 4-D by a hyperplane $w=0$. 
  The white points are cross sections of two edges of the triangle.
  The line connecting the two points are a cross section of the triangle.
  (b) A cross section of a regular tetrahedron by $w=0$. 
  The quadrilateral with white vertices is a cross section of the tetrahedron by the hyperplane.
  \label{115533}}
\end{figure}

A regular pentachoron has three regular tetrahedra around each edge.
A slice of a regular pentachoron in 4-D with a hyperplane is a 3-D object that is composed of vertices, edges, and faces in 3-D.
Here we visualize them with OpenGL.
Consider a hyperplane $w=0$.
The slice of an edge in 4-D by the hyperplane is a vertex in 3-D.
Figure~\ref{115533}(a) schematically shows a triangle in 4-D that is sliced by the hyperplane $w=0$.
The two white points are slices of two edges in the triangle.
They will be visualized by balls in 3-D with OpenGL.
The thick line connecting them is a slice of the triangle.
It will be shown as a bar in 3-D.

\begin{figure}[t]
  \centering
  \includegraphics[%
      height=0.8\textheight,%
       width=0.8\hsize,keepaspectratio]%
        {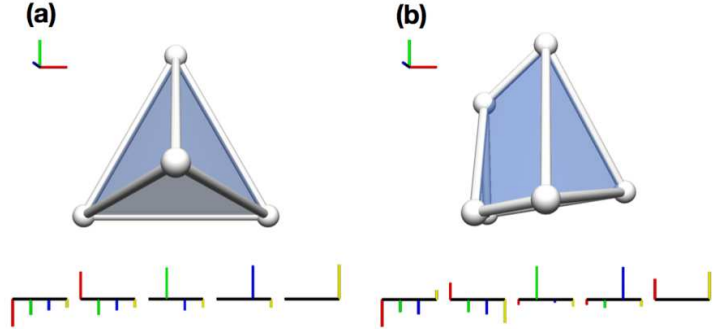}
  \caption{3-D slices of a 4-D regular pentachoron with a hyperplane. 
  A slice is composed of vertices, edges, and faces
  that are visualized by balls, bars, and semi-transparent planes.
  (a) A slice with the hyperplane $w=0$ of a regular pentachoron before applying any rotation.
The 3-D axes in the upper left depict $x$ (red), $y$ (green), and $z$ (blue) directions.
The colored vertical lines at the bottom are parallel coordinates of the five vertices of the regular pentachoron.
Details are described in the text.
  (b) A slice of the regular pentachoron by the same hyperplane after a rotation is applied.
The change of the five vertices' positions are apparent in the parallel coordinates.
  }\label{142541}
\end{figure}

A slice of a tetrahedron in the regular pentachoron is either a triangle or a quadrilateral.
An example of a quadrilateral cross section is shown in Fig.~\ref{115533}(b).
The four white points constitute a quadrilateral.
The 3-D coordinates of these four points are already calculated when the vertices (balls) and edges (bars) are visualized.
Note that the four points are on a plane in 3-D because they are on two hyperplanes in 4-D.
The quadrilateral is drawn as a semitransparent face in 3-D. 
Examples of 3-D slices of a 4-D regular pentachoron are shown in Fig.~\ref{142541}.
The panel~(a) of this figure is a 3-D slice of a regular pentachoron by $w=0$ hyperplane, viewed in the negative $z$ direction.
The RGB-colored axes in the upper left in each panel of Fig.~\ref{142541} depict the $x$, $y$, and $z$ directions, respectively.
The colored vertical lines the bottom in each panel will be explained in the next section.

\section{Multiple Slices and Their Oval Display}

\begin{figure}[t]
  \centering
  \includegraphics[%
      height=0.99\textheight,%
       width=0.99\hsize,keepaspectratio]%
        {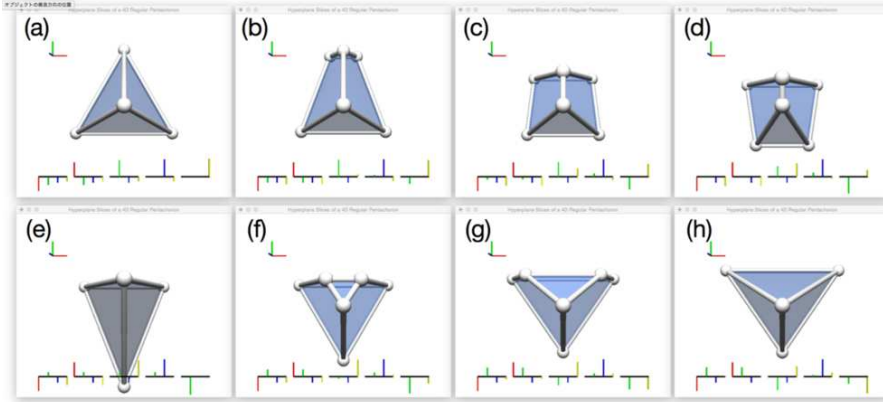}
  \caption{A sequence of a simple rotation in 4-D. The slice is a hyperplane $w=0$.
  The regular pentachoron is rotated in the $y$--$w$ plane for $\pi$ radian from (a) to (g).
  The colored parallel coordinates of the five vertices at the bottom of each panel
  indicate that $y$ coordinates (green) and $w$ coordinates (yellow) are mixed as the rotations goes on, while other coordinates, $x$ (red), $z$ (blue) remain the same.
  }\label{140157}
\end{figure}

%
The edge length of the regular pentachoron, that is presented in Fig.~\ref{142541} and other figures in this paper, is 2.
The initial condition of the regular pentachoron is shown in Fig.~\ref{142541}(a).
The coordinates $(x,y,z,w)$ of the five vertices $P_0$ to $P_4$ in the initial condition are as follows:
$P_0:(-1,-1/\sqrt{3},-1/\sqrt{6},-1/\sqrt{10})$,
$P_1:(1,-1/\sqrt{3},-1/\sqrt{6},-1/\sqrt{10})$,
$P_2:(0,2/\sqrt{3},-1/\sqrt{6},-1/\sqrt{10})$,
$P_3:(0,0,\sqrt{3}/\sqrt{2},-1/\sqrt{10})$,
and
$P_4:(0,0,0,4/\sqrt{10})$.
The center of gravity is on the origin.
The vertical short lines at the bottom of Fig.~\ref{142541}
are parallel coordinates for $x$ (red), $y$ (green), $z$  (blue) and $w$ (yellow), of the five vertices $P_0$ to $P_4$.
They are placed from left to right in the panel.
Figure~\ref{142541}(b) shows a slice with the same $w=0$ plane of the regular pentachoron that is rotated in 4-D.

A rotation in 4-D is represented by a so-called double rotation.
A double rotation is a combination of two simple rotations around two fixed planes that are absolutely perpendicular each other~\cite{cole1890rotations,Manning:1914}.
Here we denote a simple rotation in, for example, the $x$--$y$ plane with rotation angle $\alpha$ as $R_{xy}(\alpha)$.
We also denote a double rotation with angles $\alpha$ and $\beta$ in the $x$--$y$ and  $z$--$w$ planes, respectively, as $R_{xy,zw}(\alpha,\beta)$.
The rotation applied in Fig.~\ref{142541}(b) from the initial condition~(a) is double rotation $R_{xw,yz}(\alpha,\beta)$,
where $\alpha=-\pi/\sqrt{3}$ and $\beta= \pi/\sqrt{3} - \pi/\sqrt{8}$.
The coordinates of the five vertices are shown in the parallel coordinates at the bottom of the panel~(b).

To apply rotations (both simple and double) in 4-D, we use the keyboard-based interface proposed in our paper~\cite{Kageyama:JVis:2015a}.
For example, typing ``2'' key causes a single rotation $R_{xy}(\alpha)$ 
and typing ``y'' key causes a double rotation $R_{xy,zw}(\alpha,\beta)$.
The rotation angle $\alpha$ is controlled by the keyboard, too.
Typing ``k'' key (``j'' key) increases (decreases) the angle $\alpha$.
This is following the vi-editor's convention.

Figure~\ref{140157} shows a sample sequence of a continuously applied simple rotation in the $y$--$w$ plane, $R_{yw}$, for  $\pi$ radian in total.
The parallel coordinates of the five vertices indicate that $y$ coordinates (green) and $w$ coordinates (yellow) are mixed as the rotations goes on, while other coordinates, $x$ (red), $z$ (blue) remain the same.

While a single slice by a hyperplane conveys very limited information of the target's 4-D structure,
presenting multiple slices, at once, by parallel hyperplanes will facilitate a grasp of the four dimensional configuration as a whole.

Presenting many images that are taken with different parameters 
is not new in the visualization of higher dimension.
For example, HyperSlice~\cite{van1993hyperslice} is a visualization method for $N\: (>2)$ dimensional scalar field.
In their method, a point of focus $\bm{c}=(c_0, c_1, \ldots c_{N-1})$ is
placed in the $N$ dimeinsional scalar field.
A 2-D field $S_{ij}(x_i,x_j)$ is defined for each $i$ and $j$  $(0\le i \ne j \le N)$ as $f(c_0, \ldots, c_{i-1}, x_i, c_{i+1}, \ldots, c_{j-1}, x_j, c_{j+1}, \ldots, c_{N-1})$.
By applying standard visualization methods for 2-D,
each field $S_{ij}$ is converted into an image $\hat{S}_{ij}$.
In the HyperSlice method, all the images $\hat{S}_{ij}$ are placed in a matrix form.
Comparing whole images in a sight, 
it is expected that one can grasp the $N$ dimensional structure of the scalar field $f$.

\begin{figure}[t]
  \centering
  \includegraphics[%
      height=0.99\textheight,%
       width=0.99\hsize,keepaspectratio]%
        {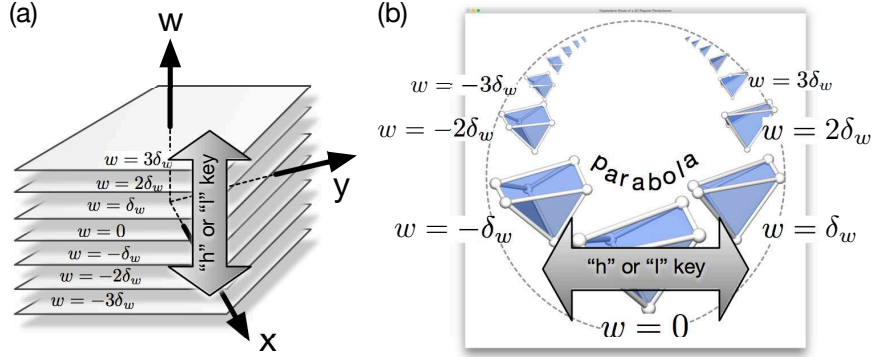}
  \caption{(a) Multiple slices in 4-D with $w=\text{const.}$ hyperplanes.
  (b) Oval display of the multiple slices.
  The 3-D slices are placed on a parabola in the world coordinates in 3-D and viewed by a perspective projection.}\label{232015}
\end{figure}

In our method, we take multiple slices of a 4-D polytope
with  hyperplanes of $w=0, \pm \delta_w, \pm 2\delta_w, \pm 3 \delta _w, \ldots$,
as schematically shown in Fig.~\ref{232015}(a).
In order to present as many slices as possible to the viewer,
we place them in a parabola curve on a plane in a 3-D world coordinates, as shown in Fig.~\ref{232015}(b).
The slices appear in a oval configuration since a parabola looks as an ellipse in the usual (3-D) projection.
In the default state, the slice by the hyperplane $w=0$ is shown at the center.
Other slices of $w=n\, \delta_w$ ($n\ge 1$) are placed in the $n$-th neighbor in the right.
Similarly, the slices of negative $w$ are aligned in the left.

The viewer can change the default value of $w$ that is shown at the center by the keyboard typing.
When the ``l'' key is typed, the default $w$ shifts to $w=+\delta_w$.
Repeated typing of the ``l'' key continuously shifts the central $w$ value.
As a consequence, in the oval display, a typing of the ``l'' key effectively shifts the ``focus slice'',
which is shown in the center, to the right neighbor.
This is again following the vi-editor's convention.
Conversely, a type of the ``h'' key causes the shift of the ``focus slice'' to the left.

\begin{figure}[t]
  \centering
  \includegraphics[%
      height=0.8\textheight,%
       width=0.8\hsize,keepaspectratio]%
        {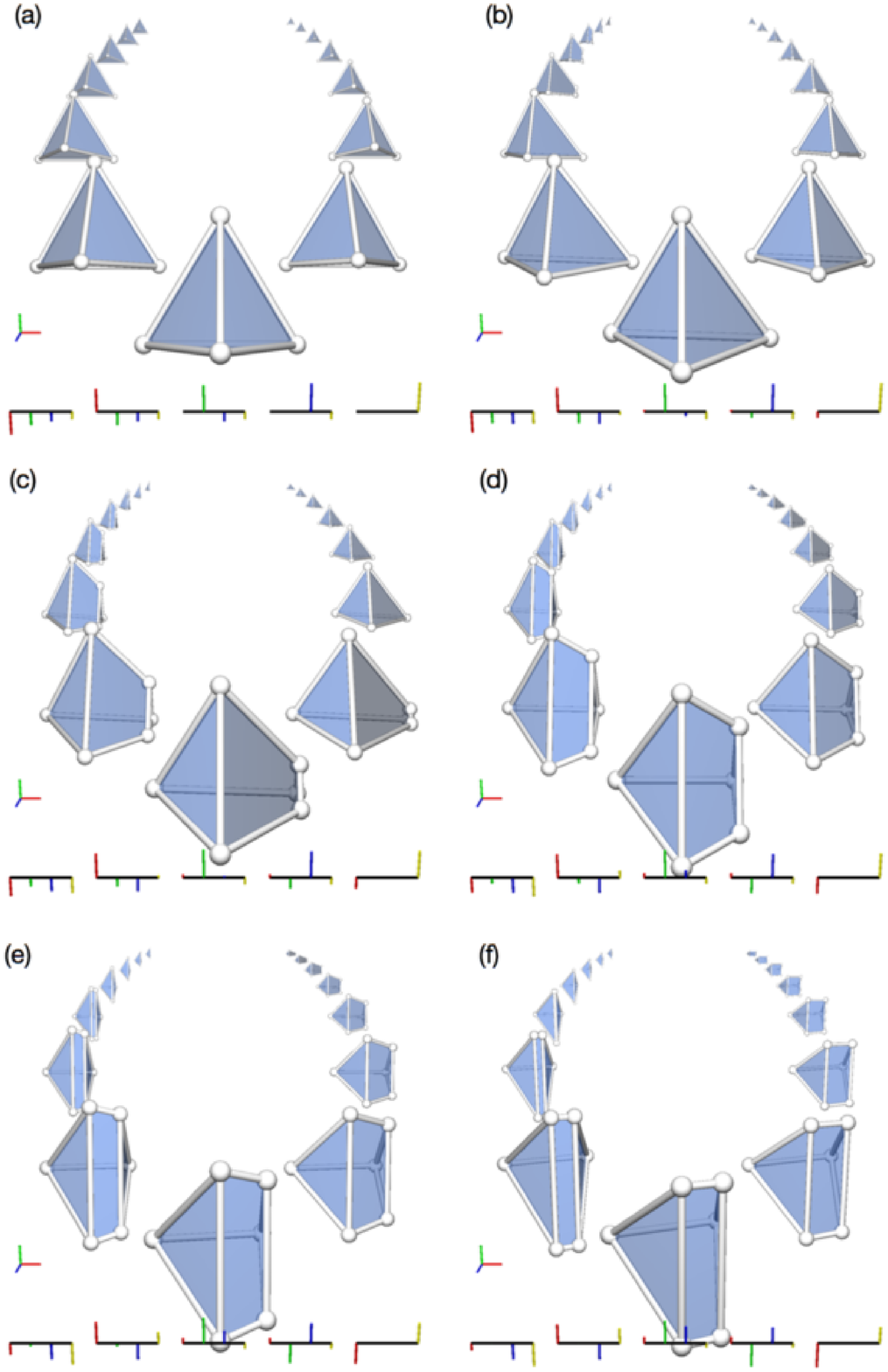}
  \caption{
  A time sequence of the oval view of a regular pentachoron.
  A double rotation $R_{xw,yz}(\alpha,\beta)$ is continuously applied to the regular pentachoron in the panels (a) to (f).
The angles are $\alpha=\pi/8$ and $\beta=\pi/\sqrt{3} - \pi/8$.
  }\label{232048}
\end{figure}

A rotation applied to the 4-D object leads to synchronous deformations of the slices.
The viewer can observe a kind of 4-D tomography of the object, interactively applying both the simple and the double
rotations in the 4-D space through the keyboard. 
An example of continuously applied 4-D rotation to a 4-D regular pentachoron is shown in Fig.~\ref{232048}.
The panels (a) to (f) show a sequence of a double rotation.
The applied rotation is $R_{xw,yz}(\alpha,\beta)$.
In the sequence of (a) to (f), the rotation is consecutively applied with fixed double 
angles $\alpha=\pi/8$ and $\beta=\pi/\sqrt{3} - \pi/8$, to move from (a) to (b), then from (b) to (c), and so on.
One can observe the $x$--$w$ and $y$--$z$ coordinates pairs are mixed in these rotations from the parallel coordinates.

%

\section{Summary}

We have proposed a visualization method of 4-D polytopes.
In the proposed method,
multiple 3-D slices of a polytope by parallel hyperplanes are presented in oval configuration in a view window.
While the hyperplanes are fixed in 4-D, 
the polytope is interactively rotated in 4-D by the viewer through the key typing of the keyboard.
Both the simple and the double rotations are applied by single key typings.
The compact view in the oval configuration of many slices facilitates a grasp of 4-D structure of the polytope. 
This visualization method is applicable to other 4-D objects than polytopes,
especially to the 4-D data visualization produced by CFD.






\end{document}